\documentclass[12pt]{article}
\usepackage{fullpage,citesort,epsfig,psfrag,graphics,amsbsy,amssymb,
}
\usepackage{caption}
\newcommand{\beq}{\begin{equation}}
\newcommand{\eeq}{\end{equation}}
\newcommand{\bea}{\begin{eqnarray}}
\newcommand{\eea}{\end{eqnarray}}
\newcommand{\bfs}{\boldsymbol}

\newcommand{\be}{\begin{equation}}
\newcommand{\ee}{\end{equation}}
\newcommand{\bq}{\begin{eqnarray}}
\newcommand{\eq}{\end{eqnarray}}
\newcommand{\ket}[1]{|#1\rangle}

\def\math{\mathsurround=0pt }
\def\leftrightarrowfill{$\math \mathord\leftarrow \mkern-6mu
 \cleaders\hbox{$\mkern-2mu \mathord- \mkern-2mu$}\hfill
 \mkern-6mu \mathord\rightarrow$}
\def\overleftrightarrow#1{\vbox{\ialign{##\crcr
     \leftrightarrowfill\crcr\noalign{\kern-1pt\nointerlineskip}
     $\hfil\displaystyle{#1}\hfil$\crcr}}}

\newcommand{\VEV}[1]{\langle#1\rangle}

\let\l=\lambda

 \def\bd{\begin{document}} \def\ed{\end{document}}
\def\ds{\documentstyle} \let\fr=\frac \let\bl=\bigl \let\br=\bigr
\let\Br=\Bigr \let\Bl=\Bigl
\let\bm=\bibitem
\let\na=\nabla
\let\pa=\partial \let\ov=\overline
\def\ft#1#2{{\textstyle{{\scriptstyle #1}\over {\scriptstyle #2}}}}
\def\fft#1#2{{#1 \over #2}}
\def\vp{\varphi}
\def\sst#1{{\scriptscriptstyle #1}}
\def\oneone{\rlap 1\mkern4mu{\rm l}}
\def\td{\tilde}
\def\wtd{\widetilde}
\def\dalemb#1#2{{\vbox{\hrule height .#2pt
        \hbox{\vrule width.#2pt height#1pt \kern#1pt
                \vrule width.#2pt}
        \hrule height.#2pt}}}
\def\square{\mathord{\dalemb{6.8}{7}\hbox{\hskip1pt}}}
\def\wtd{\widetilde}
\def\R{\rlap{\rm I}\mkern3mu{\rm R}}
\def\im{{\rm i}}
\def\tilg{\tilde{g}}
\def\tilF{\tilde{F}}
\def\tilA{\tilde{A}}
\def\varf{\varphi}
\def\tilf{\tilde{\phi}}
\def\tilh{\tilde{h}}
\def\rme{{\rm e}}
\def\ep{\epsilon}
\def\0{{(0)}}
\def\9{{(9)}}
\def\8{{(8)}}
\def\7{{(7)}}
\def\6{{(6)}}
\def\5{{(5)}}
\def\4{{(4)}}
\def\3{{(3)}}
\def\2{{(2)}}
\def\1{{(1)}}
\newcommand{\trace}{{\rm Tr}}
\newcommand{\ub}{\overline{U}}
\newcommand{\vb}{\overline{V}}
\newcommand{\uh}{\widehat{U}}
\newcommand{\vh}{\widehat{V}}
\newcommand{\ubh}{\overline{\widehat{U}}}
\newcommand{\vbh}{\overline{\widehat{V}}}
\newcommand{\lb}{\bar{\l}}
\newcommand{\Fb}{\overline{F}}
\newcommand{\Fh}{\widehat{F}}
\newcommand{\Fbh}{\overline{\widehat{F}}}
\newcommand{\Ab}{\overline{A}}
\newcommand{\Ah}{\widehat{A}}
\newcommand{\Abh}{\overline{\widehat{A}}}
\newcommand{\Gb}{\overline{G}}
\newcommand{\Gh}{\widehat{G}}
\newcommand{\Gbh}{\overline{\widehat{G}}}
\newcommand{\Pb}{\overline{P}}
\newcommand{\Ph}{\widehat{P}}
\newcommand{\Pbh}{\overline{\widehat{P}}}
\newcommand{\Qb}{\overline{Q}}
\newcommand{\Qh}{\widehat{Q}}
\newcommand{\Qbh}{\overline{\widehat{Q}}}
\newcommand{\Bb}{\overline{B}}
\newcommand{\Bh}{\widehat{B}}
\newcommand{\Bbh}{\overline{\widehat{B}}}
\newcommand{\fhns}{\hat{F}^{\rm (NS)}}
\newcommand{\fhrr}{\hat{F}^{\rm (RR)}}
\newcommand{\ahns}{\hat{A}^{\rm (NS)}}
\newcommand{\ahrr}{\hat{A}^{\rm (RR)}}
\newcommand{\hhrr}{\hat{H}^{\rm (RR)}}
\newcommand{\hchi}{\hat{\chi}}
\newcommand{\hphi}{\hat{\phi}}
\newcommand{\htau}{\hat{\tau}}
\newcommand{\cG}{{\cal G}}
\newcommand{\cGb}{\overline{{\cal G}}}
\newcommand{\cH}{{\cal H}}
\newcommand{\cP}{{\cal P}}
\newcommand{\cPb}{\overline{{\cal P}}}
\newcommand{\cQ}{{\cal Q}}
\newcommand{\cQb}{\overline{{\cal Q}}}
\newcommand{\cM}{{\cal M}}
\newcommand{\cN}{{\cal N}}
\newcommand{\cO}{{\cal O}}
\newcommand{\cD}{{\cal D}}
\newcommand{\cL}{{\cal L}}

\newcommand{\vpp}{\mbox{$\langle{\scriptstyle++}\rangle$}}
\newcommand{\vmp}{\mbox{$\langle{\scriptstyle-+}\rangle$}}
\newcommand{\vppp}{\mbox{$\langle{\scriptstyle+++}\rangle$}}
\newcommand{\vmpp}{\mbox{$\langle{\scriptstyle-++}\rangle$}}
\newcommand{\vpmp}{\mbox{$\langle{\scriptstyle+-+}\rangle$}}

\begin{document}
\setlength{\captionmargin}{36pt}
\begin{titlepage}
\begin{flushright}
\phantom{UFIFT-HEP}
\end{flushright}

\vskip 3cm
\begin{center}
\begin{large}
{\bf Worldsheet Propagator on the Lightcone Worldsheet Lattice}
\end{large}

\vskip 2cm
{\large
Georgios Papathanasiou\footnote{E-mail  address: {\tt georgios@ufl.edu}} and Charles B. Thorn\footnote{E-mail  address: {\tt thorn@phys.ufl.edu}}
}
\vskip0.20cm
{\it Institute for Fundamental Theory\\
Department of Physics, University of Florida,
Gainesville FL 32611}


\vskip 1.0cm
\end{center}

\begin{abstract}
\noindent We develop new more powerful techniques, based on an 
almost closed form for the lattice worldsheet propagator, for analyzing
planar open string worldsheets defined on a lightcone lattice.
We show that results obtained in earlier work are easily reproduced
with far more precision. In particular, consistency checks which
required numerical analysis in the earlier work can now
be confirmed exactly.
\end{abstract}
\vfill
\end{titlepage}
\section{Introduction}
The lightcone worldsheet \cite{goddardrt,goddardgrt,mandelstamlc}
lattice was proposed long ago \cite{gilest} as
a method to digitize the summation of planar
open string multiloop diagrams. Because the open string spectrum includes
a massless spin 1 particle, this sum of diagrams should have
the infrared behavior of large $N$ \cite{thooftlargen} gauge theory.
If the worldsheet lattice can reliably reproduce string theory
diagrams, its $\alpha^\prime\to0$ limit should just as
reliably reproduce gauge theory \cite{thornsubqcd}.
With this possibility in mind, we have recently embarked on a
program \cite{papathanasiout} to critically evaluate the accuracy
of this lattice in reproducing the continuum perturbative diagrams.
In particular, it is very important that lattice artifacts be
shown to be benign: they should either vanish in the continuum
limit or be absorbed in redefinitions of parameters
in the theory. This is a necessary prerequisite to applying
these lattice methods to nonperturbative calculations of QCD.

In \cite{papathanasiout} we studied the one loop self-energy diagram for the
bosonic closed string
in enough detail to see that the lattice accurately reproduced
the ultraviolet behavior of the diagram. This is as much as we should
expect, since the open bosonic string tachyon
should and does ruin the infrared behavior of the diagram\footnote{
It is logically possible that summing the bosonic string
diagrams with an infrared cutoff stabilizes the vacuum in a way to produce
QCD physics, but it is more likely that another string model, such as the
tachyon-free
Neveu-Schwarz sector of the superstring, is required to truly reproduce
QCD.}.
The analysis in \cite{papathanasiout} employed what might be called a string field theory
 approach (see, for example \cite{thornsft}): the diagram
was built up from open and closed string propagators. While this
approach was manageable at one loop, it quickly gets unwieldy
for multiloop diagrams. Even for the one loop open string self-energy
diagram, attaining enough accuracy to make definitive conclusions
proved to be problematic. To improve on this situation
we develop, in this article, a more powerful ``worldsheet''
approach based on the techniques of
worldsheet quantum field theory defined on the lightcone lattice.
The key to this approach is an almost closed form expression for
the worldsheet propagator on the lattice (see Eq.(\ref{closed_Delta})).

The perturbative string field theory approach of \cite{papathanasiout} keeps manifest the
contribution of all the intermediate string mass eigenstates
contributing to the diagram. While the ensuing formulas for
the self-energy shifts were exact at finite lattice
spacing, we had to resort
to numerical analysis to analyze the continuum limit.
The extrapolation of our numerical results to the continuum
was sufficiently accurate to
make rather convincing consistency checks, such as a vanishing graviton self-energy. These checks were nonetheless subject to numerical
error. In contrast, the methods of the present paper are powerful enough
to analyze the continuum limit exactly in the ultraviolet
and to confirm rigorously such consistency requirements.

The Giles-Thorn (GT) discretization of the worldsheet \cite{gilest}
begins with a representation of the
free closed or open string propagator as a lightcone worldsheet
path integral defined on a lattice. The lattice replaces the
transverse coordinates of the string ${\bfs x}(\sigma,\tau)$, living on
a rectangular $P^+\times T$ domain, with discretely
labeled coordinates ${\bfs x}_k^j={\bfs x}(kaT_0,ja)$, living
on an $M\times N$ grid with spacing $a$, where $P^+=MaT_0$ and $T=a(N+1)$.
The free string propagator is then simply a Gaussian integral
\bea
{\cal D}_0&=&\int \prod_{kj}d{\bfs x}_k^je^{-S},\nonumber\\
S&=&{T_0\over2}\sum_{kj}\left[({\bfs x}_k^{\ j+1}-{\bfs x}_k^{\ j})^2
+({\bfs x}_{k+1}^{\ j}-{\bfs x}_k^{\ j})^2\right]\equiv
{T_0\over2}{\bfs x}^T\cdot\Delta^{-1}{\bfs x}\,,
\eea
where the $MN\times MN$ matrix
$\Delta$ is the lattice worldsheet propagator that will
be the central focus of this article. Then up to an overall
normalization factor ${\cal D}_0={\det}^{-(D-2)}\Delta^{-1}$,
where $D$ is the spacetime dimension ($D=26$ for the bosonic string).

On this lattice the sum of all
open string multiloop planar diagrams can
be obtained by summing over all patterns of missing spatial
bonds. Formally, this is achieved by introducing Ising-like variables
$S_k^j=0,1$ and taking the worldsheet action to be
\bea\label{action_isinglike}
S_{\rm Planar}&=&{T_0\over2}\sum_{ij}\left[({\bfs x}_i^{\ j+1}
-{\bfs x}_i^{\ j})^2
+S_i^{\ j}({\bfs x}_{i+1}^{\ j}-{\bfs x}_i^{\ j})^2\right]\nonumber\\
&&+(D-2)B\sum_{kj}(1-S_k^j)-\sum_{ij}\left[S_i^{\ j}(1-S_i^{\ j+1})
+S_i^{\ j+1}(1-S_i^{\ j})\right]\ln g\\
&\equiv&{T_0\over2}{\bfs x}^T\cdot\left[\Delta^{-1}+V(S)\right]{\bfs x}
+A(\{S\})\label{lattice_action}\,.
\eea
The terms in $A(\{S\})$ insert the coupling constant $g$ in the
appropriate way and allow for an open string self-energy counterterm
$B$. Then we have
\bea
{\cal D}&=&{\cal D}_0\sum_{\{S\}}{\det}^{-12}(I+V\Delta)e^{-A(\{S\})}\,.
\eea
When $V$ is a sparse matrix, i.e. when there are a relatively small
number of missing bonds ($\sum_{kj}(1-S_{kj})\ll M$ which
can be arranged by taking $B\gg1$),
this will be a particularly efficient
way to evaluate the terms of perturbation theory. Holding $B$
sufficiently large serves as a physical and convenient infrared
regulator in our studies of the properties of the planar diagrams.

The paper is organized as follows. In Section 2 we construct the
worldsheet propagator on the GT worldsheet lattice. It is remarkable
that, in spite of the discretization of time ($ix^+$), the result is explicit
and not much more complicated than the well known continuum worldsheet
propagator. In section 3 we apply this expression to the calculation of
the tachyon one loop closed string self energy. Our results, being exact, can
be carried out to arbitrary precision, agreeing with the numerical
results of \cite{papathanasiout} to the precision achieved in 
that reference (only 3 significant
figures for some of the subleading contributions). The self-energy
of the closed string graviton and selected higher mass states
is similarly analyzed in Section 4. We conclude
with discussion in Section 5 of the significance of our
results and their promise for analyzing the open string
self-energy as well as higher loop diagrams.
\section{Lattice Worldsheet Propagators}
We develop the tools of quantum field theory for the worldsheet lattice.
Of central interest are the worldsheet correlators of the
coordinates on the $M\times N$
lattice corresponding to the free closed or open string,
\bea\label{wscorrelator_definition}
\Delta_{ij,kl}=T_0\VEV{x_i^jx_k^l}=T_0{\int {\cal D}x\ x_i^jx_k^l\ e^{-S}\over
\int {\cal D}x\ e^{-S}}\,.
\eea
Because the expectations are taken with Gaussian weight, the
two point correlator in a single dimension captures all of the
relevant information in arbitrary multi-point
correlators in any number of dimensions. For the bosonic string
we should of course take 26 space-time dimensions or 24 transverse
dimensions.

A straightforward evaluation is to use
closure to write the numerator as the product of three string
field propagators
(see Appendix~\ref{propagators}): one from time $-(N-j)$ to $j$, one
from time $j$ to $l$, and the last from time $l$ to $+(N+l)$. We can
resolve $x_i^j$,  $x_k^l$ into normal modes $q_m^j$, $q_n^l$ respectively.
Then because each normal mode path integral is independent,
$\VEV{q_m^jq_n^l}=\delta_{mn} \VEV{q_m^jq_m^l}$ one ends up with
a simple two variable Gaussian integral to do,
\bea
\int dq_m^j\ dq_m^l q_m^j q_m^l\exp\left\{-{1\over2}[A(q_m^{j2}+q_m^{l2})
+2Bq_m^j q_m^l]\right\}&=&-{B\over A^2-B^2}{\det}^{-1/2}
\pmatrix{A&B\cr B& A\cr}\nonumber\\
\VEV{q_m^jq_n^l}&=&-{B\over A^2-B^2}\delta_{mn}\,.
\eea
Here $A$ and $B$ are read off from the formulas of Appendix~\ref{propagators}.
For simplicity we set the $q$'s at the initial and final times to zero.

Then for non-zero modes they are:
\bea
A&=&T_0\sinh\lambda\left[\coth N\lambda+\coth(l-j)\lambda\right],\qquad
B={-T_0\sinh\lambda\over\sinh(l-j)\lambda}\,,
\eea
where $\lambda$ is $\lambda_m^o=2\sinh^{-1}\sin(m\pi/2M)$
or $\lambda_m^c=2\sinh^{-1}\sin(m\pi/M)$ for the open or closed
string respectively. The non-zero mode contribution has a well defined
$N\to\infty$ limit:
\bea
A&\to&T_0\sinh\lambda\left[1+\coth(l-j)\lambda\right],\qquad
B={-T_0\sinh\lambda\over\sinh(l-j)\lambda}\,,\nonumber\\
{-B\over A^2-B^2}&\to&{1\over 2T_0\sinh\lambda}\left({1\over(\cosh(l-j)\lambda
+\sinh(l-j)\lambda)}\right)={e^{-|l-j|\lambda}\over2T_0\sinh\lambda}\,.
\eea
For the zero modes
\bea
A_0&=&T_0{N+l-j\over N(l-j)},\qquad B_0=-{T_0\over l-j}\,,\nonumber\\
{-B_0\over A_0^2-B_0^2}&=&{N\over2T_0(1+(l-j)/2N)}\to {N\over2T_0}-{l-j\over4T_0}\,,
\eea
where we have taken $N$ large on the right side. To properly isolate
the diverging term, we must remember that $2N$ is not the total time
length of the string propagator. Rather the total length
is $2N_T=2N+l-j$. $N_T$ is the quantity that should be
regarded as independent of $j,l$.
In other words we should write
\bea
\VEV{q_0^jq_0^l}&\sim&{N\over2T_0}-{l-j\over4T_0}+\mathcal{O}\left(\frac{1}{N}\right)
={N_T\over2T_0}-{l-j\over2T_0}+\mathcal{O}\left(\frac{1}{N_T}\right)\,.
\eea
The zero mode contribution grows linearly with $N_T$. So it will be important
that the zero mode be suppressed in the physical quantities that
require the input of worldsheet propagators. It is helpful to appreciate
though that the divergent term is a constant independent of $j,l$.
To define the inverse lattice Laplacian, it is consistent to drop
it in effect modifying the boundary conditions on the Green function.

In these derivations we have assumed $l>j$. For $l<j$ the roles
of the two indices are switched. When $N\to\infty$, we simply
replace $l-j\to|l-j|$ in the formulas:
\bea\label{wsprop_fourier}
\VEV{q_m^jq_n^l}&=&\delta_{mn}
{e^{-|l-j|\lambda_m}\over2T_0\sinh\lambda_m},\qquad N\to\infty\,,
\nonumber\\
\VEV{q_0^jq_0^l}&\sim&{N_T\over2T_0}-{|l-j|\over2T_0},\qquad N\to\infty\,.
\eea
From their physical interpretation these are inverses of the
lattice Laplacian\footnote{Similar results for the
inverse of the discrete Laplacian (discrete Green function) 
have actually appeared some time ago \cite{discretegreen}. }
\bea(-\triangle+4\sinh^2\lambda/2)f_j
\equiv 2f_j-f_{j+1}-f_{j-1}+4f_j\sinh^2\lambda/2\,.
\eea
It is remarkable that this can be checked directly:
\bea
2e^{-|l-j|\lambda}-e^{-|l+1-j|\lambda}-e^{-|l-1-j|\lambda}&=&
\cases{e^{-(l-j)\lambda}\left(2-2\cosh\lambda\right)&$l>j$\cr
e^{-(j-l)\lambda}\left(2-2\cosh\lambda\right)&$l<j$\cr
2-e^{-\lambda}-e^{-\lambda}=(2-2\cosh\lambda)+2\sinh\lambda&$l=j$}\nonumber\\
&=&-4e^{-|l-j|}\sinh^2{\lambda\over2}+2\delta_{lj}\sinh\lambda\,,\\
\left(-\triangle+4\sinh^2{\lambda\over2}\right){e^{-|l-j|\lambda}\over
2\sinh\lambda}&=&\delta_{lj}\,,
\eea
which shows that $\VEV{q_m^jq_n^l}$ is the inverse of the
lattice Laplacian on the nonzero modes. The proof for zero modes is even
simpler
\bea
-\triangle |l-j|&=&2|l-j|-|l+1-j|-|l-1-j|=-2\delta_{lj}\,,
\eea
which confirms the same property for the zero modes.

Finally we return to the correlators on the spatial lattice by
expanding in normal modes. The mode functions differ for the
various types of string. For the Neumann open string
\bea\label{neumann_open_Delta}
\Delta^o_{hj,kl}&=&T_0\VEV{x_h^jx_k^l}={T_0\over M}\VEV{q_0^jq_0^l}
+{2T_0\over M}\sum_{m=1}^{M-1}
\VEV{q_m^jq_m^l}\cos{m(h-1/2)\pi\over M}\cos{m(k-1/2)\pi\over M}\nonumber\\
&=&{N_T-|l-j|\over 2M}+{1\over M}\sum_{m=1}^{M-1}
{e^{-|l-j|\lambda^o_m}\over\sinh\lambda^o_m}
\cos{m(h-1/2)\pi\over M}\cos{m(k-1/2)\pi\over M}\; .
\eea
For the Dirichlet open string
\bea
\Delta^D_{hj,kl}&=&T_0\VEV{y_h^jy_k^l}={2T_0\over M}\sum_{m=1}^{M-1}
\VEV{q_m^jq_m^l}\sin{mh\pi\over M}\sin{mk\pi\over M}\nonumber\\
&=&{1\over M}\sum_{m=1}^{M-1}
{e^{-|l-j|\lambda^D_m}\over\sinh\lambda^D_m}
\sin{mh\pi\over M}\sin{mk\pi\over M}, \qquad h,k\neq M\nonumber\\
\Delta^D_{Mj,Ml}&=&T_0\VEV{y_M^jy_M^l}=
{e^{-|l-j|\lambda^D_M}\over2\sinh\lambda^D_M}\qquad
\Delta^D_{Mj,kl}=0,\quad k\neq M\; .
\eea
For the closed string
\bea\label{closed_Delta}
\Delta^c_{hj,kl}&=&T_0\VEV{x_h^jx_k^l}={T_0\over M}\VEV{q_0^jq_0^l}+{T_0\over M}\sum_{m=1}^{M-1}
\VEV{A_m^jA_{M-m}^l}\exp{2m(h-k)i\pi\over M}\nonumber\\
&=&{N_T-|l-j|\over 2M}+{1\over 2M}\sum_{m=1}^{M-1}
{e^{-|l-j|\lambda^c_m}\over\sinh\lambda^c_m}
\exp{2m(h-k)i\pi\over M}\,.
\label{wspropc}
\eea
\section{Closed String Self-Energy: Tachyon}
For the rest of the paper, we will apply our new approach in 
order to assess its calculational efficiency. In particular, 
we will use it to obtain 1-loop self-energy corrections to 
low-lying states of the closed string, 
for which we have a measure of comparison from our previous 
treatment \cite{papathanasiout}. In this section, 
we will particularly focus on the tachyon ground state.
\subsection{A Single Missing Link}
The matrix $V$ has indices that are lattice locations, i.e. they are
specified by two integers $V_{kj;ml}$. For a single missing link,
at time $j$ and linking spatial site $k$ to site $k+1$, the
term $(T_0/2)({\bfs x}_{k+1}^j-{\bfs x}_k^j)^2$ is missing from
$S$. That means that
\bea
\sum_{ml;m^\prime l^\prime}{\bfs x}_m^l\cdot
V_{m^\prime l^\prime;ml}{\bfs x}_{m^\prime}^{l^\prime}
=-({\bfs x}_{k+1}^j-{\bfs x}_k^j)^2=-{\bfs x}_{k+1}^{j2}
-{\bfs x}_k^{j2}+2{\bfs x}_{k+1}^j\cdot{\bfs x}_k^j\,,
\eea
from which we see
\bea\label{Vmatrix}
V_{ml;m^\prime l^\prime}&=&-\delta_{lj}\delta_{l^\prime j}
(\delta_{m,k+1}\delta_{m^\prime,k+1}+\delta_{m,k}\delta_{m^\prime,k}
-\delta_{m,k+1}\delta_{m^\prime k}-\delta_{m^\prime,k+1}\delta_{m k})\,.
\eea
This matrix has entries only in rows and columns with labels
$kj$ and $k+1,j$, in other words a $2\times2$ submatrix. However
the product matrix $V\Delta$ has nonzero entries only in rows
with labels $kj$ and $k+1,j$, but in general any column entry in these
two rows can be nonzero: there are $2MN$ entries! But in calculating
the determinant of $I+V\Delta$ by expanding in minors, one quickly sees that
it is only the $2\times2$ subblock of $I+V\Delta$ that contributes.
Similarly if there are several missing links, the only part of
$I+V\Delta$ that contributes to the determinant is a correspondingly
sized subblock.

Let us work out $V\Delta$ and  the determinant for a single
missing link,
\bea\label{VDelta_matrix}
(V\Delta)_{ml,qp}&=&-\delta_{lj}
(\delta_{m,k+1}\Delta_{(k+1)j,qp}+\delta_{m,k}\Delta_{kj,qp}
-\delta_{m,k+1}\Delta_{kj,qp}-\delta_{m k}\Delta_{(k+1)j,qp})\nonumber\\
&=&-\delta_{lj}
((\delta_{m,k+1}-\delta_{mk})(\Delta_{(k+1)j,qp}-\Delta_{kj,qp}))\,.
\eea
Then the desired determinant is
\bea
\det(I+V\Delta)&=&\det\pmatrix{1+\Delta_{(k+1)j,kj}-\Delta_{kj,kj}
&\Delta_{(k+1)j,(k+1)j}-\Delta_{kj,(k+1)j}\cr
-\Delta_{(k+1)j,kj}+\Delta_{kj,kj}&1-\Delta_{(k+1)j,(k+1)j}+\Delta_{kj,(k+1)j}}
\nonumber\\
&=&1-\Delta_{(k+1)j,(k+1)j}+\Delta_{kj,(k+1)j}+\Delta_{(k+1)j,kj}
-\Delta_{kj,kj}\,.
\eea
From (\ref{wspropc})
\bea
\Delta_{(k+1)j,kj}
-\Delta_{kj,kj}&=&{1\over 2M}\sum_{m=1}^{M-1}
{1\over\sinh\lambda^c_m}
\left(\exp{2mi\pi\over M}-1\right)\nonumber\\
&=&-{1\over M}\sum_{m=1}^{M-1}
{\sin^2(m\pi/M)\over\sinh\lambda^c_m}=-{1\over 2M}\sum_{m=1}^{M-1}
{\sin(m\pi/M)\over\sqrt{1+\sin^2(m\pi/M)}}\,.
\eea
Evidently the same result is obtained for the difference
$\Delta_{kj,(k+1)j}
-\Delta_{(k+1)j,(k+1)j}$.
For large $M$ we can apply the Euler-Maclaurin series
\bea\label{EulerMcLaurin}
{1\over M}\sum_{m=1}^{M-1}f\left({m\over M}\right)
&=&\int_0^1 dx f(x)-{1\over2M}(f(0)+f(1))\nonumber\\
&&+\sum_{k=1}^\infty{B_{2k}\over(2k)!}
{1\over M^{2k}}(f^{(2k-1)}(1)-f^{(2k-1)}(0))\,,
\eea
where $B_k$ are the Bernoulli numbers, to get
\bea
\Delta_{(k+1)j,kj}-\Delta_{kj,kj}&=&-{1\over2}\left[\int_0^1dx{\sin\pi x\over\sqrt{1+\sin^2\pi x}}-2\pi{B_2\over2M^2}
+8\pi^3{B_4\over24M^4}+{\cal O}\left({1\over M^6}\right)\right]\nonumber\\
&=&-{1\over2}\left[{1\over2}
-{\pi\over6M^2}-{\pi^3\over90M^4}+{\cal O}\left({1\over M^6}\right)\right]\,.
\eea
So finally
\bea
\det(I+V\Delta)&=&{1\over2}+{\pi\over6M^2}+\frac{\pi^3}{90M^4}+{\cal O}\left({1\over M^6}\right)\,.\eea
\subsection{Single Slit with $K-1$ Missing Links}\label{sec_longslit}
\begin{figure}
\begin{center}
\includegraphics[width=4in]{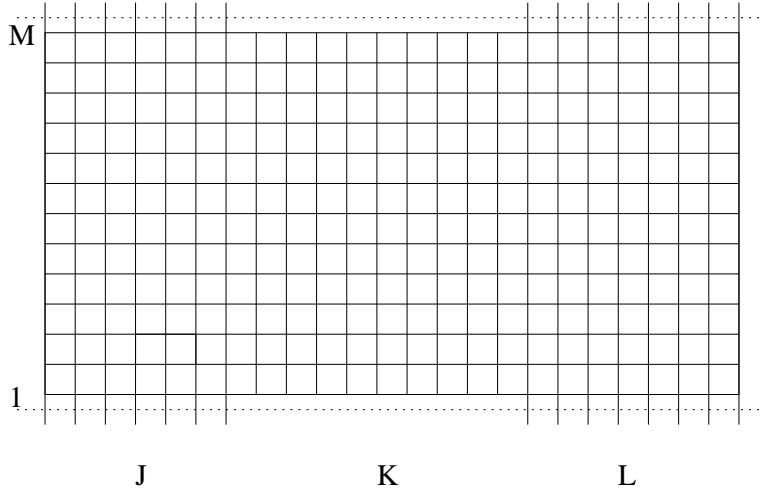}
\caption{GT worldsheet lattice for the closed string self-energy. The dotted
lines are identified. There are $K-1$ missing links, chosen for 
concreteness between the spatial positions $k=1$ and $k=M$.}
\label{closedselattice}
\end{center}
\end{figure}
The case of one missing link describes a one loop diagram
with the loop occupying two time steps. A single loop occupying
$K$ time steps has $K-1$ consecutive missing links, as we depict in figure \ref{closedselattice}.
Proceeding with the case of $K-1$ missing links,
again between spatial sites $k$ and $k+1$, but this time for
the time interval between instants $J+1$ to $J+K-1$, it is evident that we will simply have to sum the right-hand side of (\ref{Vmatrix}) over $j\in [J+1,J+K-1]$. The sum will also carry over to (\ref{VDelta_matrix}), whose nontrivial subblock will now have size $2(K-1)$. Due to the difference of delta functions on the latter relation, clearly the matrix rows with $m=k+1$ will have the opposite values of the rows with $m=k$. So sorting our rows such that
\be
(m,l)=\{(k,J+1),\ldots(k,J+K-1),\ldots,(k+1,J+1),\ldots,(k,J+K-1)\}\,,
\ee
and similarly for the columns, we can write in $(K-1)\times (K-1)$ block form
\bea\label{IVDelta_manipulate}
\det(I+V\Delta)&=&\det\pmatrix{I+A&B\cr -A& I-B\cr}=\det\pmatrix{I&I\cr -A& I-B\cr}\nonumber\\
&=&\det\pmatrix{I&0\cr -A& I+A-B\cr}=\det(I+A-B)\,,
\eea
where we employed elementary row and column manipulations that leave the determinant invariant, together with the block matrix identity
\be
\det \left(
\begin{array}{cccc}
Q& 0\\
R& S
\end{array}\right)=\det(Q) \det(S)\,.
\ee
In formula (\ref{IVDelta_manipulate}) above,
\be
A_{lp}=\Delta_{(k+1)l,kp}-\Delta_{kl,kp}\,,\qquad B_{lp}=\Delta_{(k+1)l,(k+1)p}-\Delta_{kl,(k+1)p}\,,
\ee
and since these quantities depend on $l,p$ only through $|l-p|$, the value of $J+1$ will be immaterial and we can set it to zero. Hence our final expression for the determinant will be
\be\label{h_determinant}
\det(I+V\Delta)=\det( h_{lp} )\,,\quad l,p=1,2,\ldots K-1\,,
\ee
where
\be\label{h_elements}
h_{lp}=\delta_{lp}+\Delta_{(k+1)l,kp}-\Delta_{kl,kp}+\Delta_{kl,(k+1)p}-\Delta_{(k+1)l,(k+1)p}\,.
\ee
\begin{table}
\centering
\begin{tabular}{|c|c|}\hline
$K$&$\det( h_{lp}(x) )$\\\hline
2&$\frac{1}{2}+x$\\
3&$-\frac{4}{\pi ^2}+\frac{2}{\pi }+\frac{4 x}{\pi }$\\
4&$-2-\frac{64}{\pi ^3}+\frac{16}{\pi ^2}+\frac{8}{\pi }+\left(-4+\frac{16}{\pi }\right) x$\\
5&$-16-\frac{8192}{9 \pi ^4}-\frac{2048}{9 \pi ^3}+\frac{256}{\pi ^2}+\frac{64}{3 \pi }+\left(-64-\frac{16384}{9 \pi^3}+\frac{2048}{3 \pi ^2}+\frac{512}{3 \pi }\right) x$\\
6&$-128-\frac{1441792}{81 \pi ^4}+\frac{45056}{27 \pi ^3}+\frac{3072}{\pi ^2}-\frac{512}{3 \pi }+\left(-768-\frac{8388608}{81 \pi ^4}+\frac{262144}{27 \pi ^3}+\frac{163840}{9 \pi ^2}-\frac{1024}{\pi }\right) x$\\\hline
\end{tabular}
\caption{\small Asymptotic expansion up to $\mathcal{O}(x)$, $x=\frac{\pi}{6M^2}$, of the determinant (\ref{det_h_asymptotic}), entering the tachyon energy shift (\ref{deltaP_asymptotic}), for a slit of length $K-1$, $K=2,\ldots,6$. Evidently, the coefficients of the expansion can be calculated exactly.}
\label{tab:det_h}
\end{table}
Notice in particular that we now have the determinant of a $(K-1)$-dimensional matrix, whose elements depend on differences of propagators, such that the zero modes in (\ref{neumann_open_Delta})-(\ref{closed_Delta}) always cancel out.
Specializing to the case of the closed string propagator (\ref{closed_Delta}), it is easy to show that\footnote{Evidently, $h_{lp}=h(|l-p|)$, namely $\det( h_{lp} )$ is the determinant of symmetric Toeplitz matrix. It can be shown \cite{determinants} that it further reduces to a product of two determinants of approximately half the size.}
\be\label{h_exact}
h_{lp}=\delta_{lp}-\frac{1}{M}\sum_{m=1}^{M-1}
{\sin(m\pi/M)\over\sqrt{1+\sin^2(m\pi/M)}}\left(\sin(m\pi/M)+\sqrt{1+\sin^2(m\pi/M)}\right)^{-2 |l-p|}\,,
\ee
and applying again the Euler-Maclaurin formula (\ref{EulerMcLaurin}), we obtain
\be\label{h_asymptotic}
h_{lp}=\delta_{lp}-I_{|l-p|}+\frac{\pi }{6 M^2}-\frac{(-1 + 3 |l-p|^2) \pi^3}{90 M^4}+\mathcal{O}\left(\frac{1}{M^6}\right)\,,
\ee
where we have reduced the integral $I_{|l-p|}$ to a simple finite sum in appendix \ref{appx_EulerMcLaurin}. Reexpressing
\be
h_{lp}(x)=h_{lp}(0)+x+\mathcal{O}\left(x^2\right)\,,\quad x=\frac{\pi }{6 M^2}\,,
\ee
we can separate the $M$-dependence of the determinant,
\be\label{det_h_asymptotic}
\det( h_{lp} )=\det( h_{lp}(0) )+\frac{\pi }{6 M^2}\frac{\partial}{\partial x}\left.\det( h_{lp}(x) )\right|_{x=0}+\mathcal{O}\left(\frac{1}{M^4}\right)\,.
\ee
Finally, the summand of the energy shift will be given by
\begin{eqnarray}
\delta P^-_{G,closed}&=&-\frac{e^{-24(K-1)B_0}}{\det^{12}(I+V\Delta)}=-\frac{e^{-24(K-1)B_0}}{\det^{12}(h_{lp})}\label{deltaP_exact}\\
&=&-\frac{e^{-24(K-1)B_0}}{\det^{12}( h_{lp}(0) )}\left(1-\frac{2\pi}{M^2}\frac{\frac{\partial}{\partial x}\left.\det( h_{lp}(x) )\right|_{x=0}}{\det( h_{lp}(0))}\right)+\mathcal{O}\left(\frac{1}{M^4}\right)\,,\label{deltaP_asymptotic}
\end{eqnarray}
where $h_{lp}$ is given exactly in (\ref{h_exact}) and asymptotically in (\ref{h_asymptotic}), and similarly we have expressed the summand in its exact (\ref{deltaP_exact}) and asymptotic (\ref{deltaP_asymptotic}) form.

\begin{table}
\centering
\begin{tabular}{|c|c|c|}\hline
$K$&$-\delta P^-_{G,closed} \textrm{ fit}$&$-\delta P^-_{G,closed} \textrm{ actual}$\\\hline
2&$0.1044844648\, -{1.31291}/{M^2}$&$0.104484465146 -1.31299/{M^2}$\\
3&$0.027700432\, -{0.9578}/{M^2}$&$0.0277004334342 -{0.957933}/{M^2}$\\
4&$0.010959556\, -{0.7268}/{M^2}$&$0.0109595576932 -{0.727031}/{M^2}$\\
5&$0.005388196\, -{0.5811}/{M^2}$&$0.00538819758183 -{0.581471}/{M^2}$\\
6&$0.003032942-0.4828/M^2$&$0.00303294412639 - 0.483277/M^2$\\\hline
\end{tabular}
\caption{\small Asymptotic expansion up to $\mathcal{O}(1/{M^2})$, of the tachyon energy shift summand (\ref{deltaP_asymptotic}) for $K=2,\ldots,6$, where $K-1$ is the slit length. The LHS coefficients were determined by fitting the $M$ dependence, as in \cite{papathanasiout}, with an error estimate at the order of the last digit. The RHS coefficients have been calculated exactly with the methods of the present paper, and evaluated up to the desired precision.}
\label{tab:tachyon_summand}
\end{table}

Let us now discuss how our current, worldsheet-based approach, compares to the string field theory-related approach we employed in \cite{papathanasiout}, when it comes to computing the tachyon energy shift. Clearly, the formulas we have derived here involve determinants of size roughly $K$, so that they are advantageous for analyzing the summand in the ultraviolet region $K\ll M$. Conversely, the approach \cite{papathanasiout} yields determinants of size $M$, more suitable for the infrared $K\gg M$ regime.

The fact that each approach is more suitable for one of the two domains, is also evident in our ability to derive asymptotic formulas there. Because in our earlier paper we had no such formula for analyzing the $K\ll M$ behavior of the integrand, we had to use the exact formula for the summand, evaluate it for a range of $M$ and $K$, and perform fits in both variables in order to find the respective dependence. In contrast, here we obtain explicitly the form of the asymptotic expansion in $M$, and we only need to fit for the dependence of the coefficients on $K$.

Furthermore, it is possible to compute these coefficients exactly for specific $K$, or evaluate them at arbitrary precision. As an example, we present the exact values for the first two coefficients of $\det( h_{lp}(x) )$ for $K=2,\ldots,6$ in table \ref{tab:det_h}. In table \ref{tab:tachyon_summand}, we also present the respective asymptotic expansion for the summand (\ref{deltaP_asymptotic}), and compare the values for the coefficients, on the one hand obtained with our current method, and on the other hand by fitting the $M$ dependence along the lines of \cite{papathanasiout}\footnote{See also figure 5 in the latter reference.}.

Clearly, the two results agree within our margins of error, notice however that their difference increases with $K$. This is a result of the systematic error coming from not taking into account the $\mathcal{O}(1/M^4)$ term in the fits, whose relative size also increases with $K$. Apart from the asymptotic expansion, we also confirmed that the exact formulas for the summand of the tachyon energy shift in the two approaches, (\ref{deltaP_exact}) here and (51) in \cite{papathanasiout}, agree for a large set of $M,K$ values.

\section{Closed String Self-Energy: Graviton}
To extract information, e.g. energy shifts,
about excited closed string states, we will need to consider
the propagator on a worldsheet which includes interactions.
If we denote the matrix describing a particular configuration of
missing links by $V$, as we did in (\ref{lattice_action}),
then the propagator in question will be given by
\bea\label{propagator_interacting}
\Delta^V&=&(\Delta^{-1}+V)^{-1}=\Delta(I+V\Delta)^{-1}
=\Delta-\Delta(I+V\Delta)^{-1}V\Delta\equiv\Delta-\Delta{\cal V}\Delta\,.
\eea
The final form on the right is useful when $V$ is sparse, because then
the inverse matrix appearing in the second term can be evaluated
as the inverse of the submatrix obtained by projecting onto the
sparse subspace.

Using index notation for the propagator, $\Delta^V_{kj,pq}$, and choosing the times $q$ and $j$ much earlier and much later than all of the times
occupied by $V$ respectively, we can also write
\bea
\Delta^V_{kj,pq}&=&\sum_{m,m^\prime}\exp\left\{
-j\lambda_m+q\lambda_{m^\prime}+{2\pi i\over M}(mk-m^\prime p)\right\}
{\tilde{\Delta}^V_{mm^\prime}\over2M\sqrt{\sinh\lambda_m\sinh\lambda_{m^\prime}}}\,,\\
\tilde{\Delta}^V_{mm^\prime}&=&{\delta_{mm^\prime}}
-{\tilde{\cal V}_{mm^\prime}\over2M
\sqrt{\sinh\lambda_m\sinh\lambda_{m^\prime}}}\,,\\
\tilde{\cal V}_{mm^\prime}&=&\sum_{kl,rs}\exp\left\{
l\lambda_m-s\lambda_{m^\prime}-{2\pi i\over M}(mk-m^\prime r)\right\}
{\cal V}_{kl,rs}\,.\label{calVm}\eea
We have normalized $\tilde{\Delta}^V_{mm^\prime}$ so that it is
$\delta_{mm^\prime}$ for ${\cal V}=0$.
Thus $\tilde{\Delta}^V_{mm^\prime}\det^{-12}(I+V\Delta)$ gives the
probability amplitude that the mode $m^\prime$ at early times
evolves to mode $m$ at late times. For the graviton self-energy, the relevant
process is modes $m=1, M-1$ at early times evolving to the same
modes at late times. Thus this contribution to the graviton
self energy is
\bea
-\left(\tilde{\Delta}^V_{11}\tilde{\Delta}^V_{(M-1)(M-1)}
+\tilde{\Delta}^V_{1(M-1)}\tilde{\Delta}^V_{(M-1)1}\right){\det}^{-12}
(I+V\Delta)\,.
\eea

\subsection{A Single Missing Link}
For starters, let's take $V$ with a single missing link. Its nonvanishing
$2\times2$ subblock is the matrix
\bea
V&=&\pmatrix{-1&1\cr1&-1\cr}.
\eea
Putting $A=\Delta_{(k+1)j,kj}-\Delta_{kj,kj}=\Delta_{(k+1)j,kj}
-\Delta_{(k+1)j,(k+1)j}$ the matrix $I+V\Delta$ projected onto the
subspace of $V$ and its inverse times $V$ are
\bea
I+V\Delta&=&\pmatrix{1+A&-A\cr -A&1+A\cr}\,,\nonumber\\
 {\cal V}=(I+V\Delta)^{-1}V&=&
{1\over1+2A}\pmatrix{1+A&A\cr A&1+A\cr}\pmatrix{-1&1\cr1&-1\cr}
={V\over1+2A}\,.\eea
Then we easily compute
\bea
\tilde{\cal V}_{mm^\prime}&=&-4e^{j(\lambda_m-\lambda_{m^\prime})+\pi i(m^\prime-m)(2k+1)/M}{\sin(\pi m/M)
\sin(\pi m^\prime/M)\over1+2A}\,,\\
\tilde{\cal V}_{mm}&=&-4{\sin^2(\pi m/M)\over 1+2A},\qquad
\tilde{\cal V}_{m(M-m)}=4e^{-2\pi im(2k+1)/M}{\sin^2(\pi m/M)\over 1+2A}\,.
\eea
Then we find
\bea
\tilde\Delta^V_{mm}=\tilde\Delta^V_{(M-m)(M-m)}&=&1+4{\sin^2(\pi m/M)\over2M(1+2A)\sinh\lambda_m}\nonumber\\
&=&1+\frac{2 \pi  m}{M^2}-\frac{2 m \pi ^2 \left(1+2 m^2 \pi \right)}{3 M^4}+\mathcal{O}\left(\frac{1}{M^6}\right)\,,\\
\tilde\Delta^V_{m(M-m)}&=&-4{\sin^2(\pi m/M)\over2M(1+2A)\sinh\lambda_m}e^{-2\pi im(2k+1)/M}\nonumber\\
&=&-{2\pi m\over M^2}e^{-2\pi im(2k+1)/M}+{\cal O}\left({1\over M^4}\right)\,.
\eea
The one missing link contribution to the self energy of the closed
string state $\ket{m,M-m}\pm\ket{M-m,m}$ is up to $\mathcal{O}(1/M^4)$
\bea
&&\hskip-18pt-{\tilde\Delta^V_{mm}\tilde\Delta^V_{(M-m)(M-m)}\pm\tilde\Delta^V_{m(M-m)}
\tilde\Delta^V_{(M-m)m}\over{\det}^{12}(1+V\Delta)}\nonumber\\
&&\quad \sim-\frac{\left(1+{4\pi m/ M^2}-{4 m \pi ^2 \left(1-6 m+2 m^2 \pi \right)}/{3 M^4}\right)}{\left({1/2}
+{\pi/6M^2}+{\pi^3}/{90M^4}\right)^{12}}\nonumber\\
&&\quad \sim{-2^{12}\left(1+{4\pi(m-1)\over M^2}\right.}-{ \left.\frac{2 \pi ^2 (-65+130 m-30(1\pm1) m^2+2 \pi +20 m^3 \pi )}{15 M^4}\right)}.\label{grav_shift}
\eea
This formula includes the shift for tachyon ($m=0$) and the graviton
($m=1$). The latter receives no $1/M^2$ correction, consistent with
zero shift in the continuum limit. Note also that we have assumed in
these formulas that the polarizations of the first and second
entries of $\ket{m,m^\prime}$ are different, so they don't properly
describe the dilaton self-energy shift.

Furthermore, for the graviton ($m=1$ and plus sign), we can also compare the above formula with the fits we obtained for the value of the graviton energy shift in \cite{papathanasiout}. Multiplying (\ref{grav_shift}) with the boundary counterterm $\exp(-24B_0)$, and writing the result in the notation of the latter paper, we have
\begin{eqnarray}
-\frac{1}{2}(1+C_G^K)&=&-\left(\frac{2}{1+\sqrt{2}}\right)^{12}\left(1-\frac{2 \pi ^2 (5+22 \pi )}{15 M^4}\right)+\mathcal{O}\left(\frac{1}{M^6}\right)\nonumber\\
&\simeq&-0.104484+\frac{10.1905}{M^4}+\mathcal{O}\left(\frac{1}{M^6}\right)\,.
\end{eqnarray}
The result above is in excellent agreement with the fit presented in figure 11 of \cite{papathanasiout}.

\subsection{Single Slit with $K-1$ Missing Links}\label{sec_longslit_graviton}
Let us now generalize the discussion of the previous section, for the worldsheet configuration where a link is missing between the same spatial sites $k$ and $k+1$, but for a time interval $K-1$ links long. Using the same reasoning as for the tachyon in the same configuration, it is possible to show that the matrix $\mathcal{V}$ defined in (\ref{propagator_interacting}) has the special structure
\be
\mathcal{V}_{kl,ks}=\mathcal{V}_{(k+1)l,(k+1)s}=-\mathcal{V}_{(k+1)l,ks}=-\mathcal{V}_{k+l,(k+1)s}=-h^{-1}_{ls}\,,
\ee
where $h$ is the same $(K-1)$-dimensional matrix appearing in (\ref{h_determinant}). With the help of these relations, we do the $k, r$ summation in (\ref{calVm})
\[
\tilde{\cal V}_{mm^\prime}=-\sum_{ls}e^{
l\lambda_m-s\lambda_{m^\prime}}\left(e^{{2\pi i\over M}(m^\prime-m)k}+e^{{2\pi i\over M}(m^\prime-m)(k+1)}-e^{{2\pi i\over M}(m^\prime+m^\prime k-m k)}-e^{{2\pi i\over M}(m^\prime-m k-m)}\right)h^{-1}_{ls},
\]
and if we take out an overall factor $\exp[\pi i (m^\prime-m)(2k+1)]$, this simplifies to
\be
\tilde{\cal V}_{mm^\prime}=-4e^{\pi i (m^\prime-m)(2k+1)}\sin \frac{m\pi}{M}\sin \frac{m^\prime\pi}{M}\sum_{ls}e^{
l\lambda_m-s\lambda_{m^\prime}}h^{-1}_{ls}\,.
\ee
It is evident that the above relation implies
\be
\tilde{\cal V}_{m(M-m)}\tilde{\cal V}_{(M-m)m}=\tilde{\cal V}_{mm}^2\,,\qquad\tilde{\cal V}_{(M-m)(M-m)}=\tilde{\cal V}_{mm}\,,
\ee
so that our final working formula for the graviton summand, also including the required boundary counterterm, will be
\begin{eqnarray}
\delta P^-_{\rm{Graviton}}&=&-\frac{e^{-24(K-1)B_0}}{\det^{12}(I+V\Delta)}\left(\tilde{\Delta}^V_{11}\tilde{\Delta}^V_{(M-1)(M-1)}
+\tilde{\Delta}^V_{1(M-1)}\tilde{\Delta}^V_{(M-1)1}\right)\nonumber\\
&=&-\frac{e^{-24(K-1)B_0}}{\det^{12}(I+V\Delta)}\left[\left(1
+{\tilde{\cal V}_{11}\over2M
\sinh\lambda_1}\right)^2+\left({\tilde{\cal V}_{11}\over2M
\sinh\lambda_1}\right)^2\right]\label{grav_shift_slit}\\
&=&-\frac{e^{-24(K-1)B_0}}{\det^{12}(I+V\Delta)}\left(1+2\tilde{U}+2\tilde{U}^2\right)\,,\nonumber
\end{eqnarray}
where
\be\label{Utilde}
\tilde{U}={\sin\frac{\pi}{M}\over{M
\sqrt{1+\sin^2\frac{\pi}{M}}}}\sum_{l,s=1}^{K-1}\left(\sin \frac{\pi }{M}+\sqrt{1+\sin ^2\frac{\pi }{M}}\right)^{2(l-s)}h^{-1}_{ls}\,,
\ee
and $h^{-1}$ is the inverse of the $(K-1)$-dimensional matrix with elements (\ref{h_elements}). The same interesting phenomenon that we encountered for the tachyon also appears here, namely we can reduce the size of the matrices entering the energy shift by a half.
\begin{table}
\centering
\begin{tabular}{|c|c|c|}\hline
$K$&$-\delta P^-_{\rm{Graviton}} \textrm{ fit}$&$-\delta P^-_{\rm{Graviton}} \textrm{ actual}$\\\hline
2&$0.104484465145 - 10.19/M^4$&$0.10448446514630 - 10.1905/M^4$\\
3&$0.027700433434 \
- 3.85/M^4$&$0.02770043343416 - 3.8499/M^4$\\
4&$0.010959557693 + \
1.87/M^4$&$ 0.01095955769317 + 1.8837/M^4$\\
5&$0.0053881975 + \
6.82/M^4$&$0.00538819758183 + 6.8571/M^4$\\
6&$0.003032944127 + \
11.28/M^4$&$0.00303294412639 + 11.3355/M^4$\\\hline
\end{tabular}
\caption{\small Asymptotic expansion up to $\mathcal{O}(1/{M^4})$, of the graviton energy shift summand for $K=2,\ldots,6$, where $K-1$ is the slit length. The LHS coefficients have been determined by fitting $M$, as in \cite{papathanasiout}, with an error estimate at the order of the last digit. The RHS coefficients have been calculated exactly with the methods of the present paper, and evaluated with two additional digits of precision. }
\label{tab:graviton}
\end{table}
The asymptotic expansion of (\ref{grav_shift_slit}) in $M$ readily follows from the respective expansion of $h_{ls}$ (\ref{h_asymptotic}), and in particular it is easy to show that
\be\label{h_inverse}
h^{-1}_{ls}=h^{-1}_{ls}(0)-\frac{\pi}{6M^2}\left(\sum_{i=1}^{K-1}h^{-1}_{li}(0)\right)\left(\sum_{i=1}^{K-1}h^{-1}_{is}(0)\right)+\mathcal{O}\left(\frac{1}{M^4}\right)\,,
\ee
where $h(0)$ is the $M$-independent part of the matrix $h$. Because the overall factor in (\ref{Utilde}) starts as $\mathcal{O}(1/M^2)$, we do not need additional terms in order to obtain (\ref{grav_shift_slit}) at $\mathcal{O}(1/M^4)$. In fact, if we only focus at $\mathcal{O}(1/M^2)$ for a moment, the term on the right-hand side of (\ref{grav_shift_slit}) simplifies to
\be\label{grav_factor}
1+2\tilde{U}+2\tilde{U}^2\simeq1+\frac{2\pi}{M^2}\sum_{l,s=1}^{K-1}h^{-1}_{ls}(0)=1+\frac{2\pi}{M^2}\frac{\frac{\partial}{\partial x}\left.\det( h_{lp}(x) )\right|_{x=0}}{\det( h_{lp}(0))}\,,
\ee
where for the last equality we used the identity
\be
\frac{\partial}{\partial x}\det(h)=\det(h){\rm Tr}\left(h^{-1}\frac{\partial h}{\partial x}\right)\,,
\ee
and also the fact that in our case the derivative matrix has all entries equal to one.

Comparing (\ref{grav_factor}) and (\ref{grav_shift_slit}) with (\ref{deltaP_asymptotic}), we observe that we have rigorously proven two important facts: That the leading order of the asymptotic $M$ expansion for the graviton is equal to the tachyon one, and that the subleading term is always $\mathcal{O}(1/M^4)$ for any $K\ll M$. Of course these properties were expected to hold on physical grounds, however in the approach of our previous paper, we could only obtain empirical indications about them from the fits. Finally, for sample slit lengths, we compare the coefficients of the aforementioned fits with the exact values obtained with our new method, and evaluated at higher precision, in table \ref{tab:graviton}.

\section{Discussion and Conclusion}
In this paper, we continued our investigation of lattice-regularized string theory in the lightcone gauge, by introducing a new approach for evaluating the corresponding path integral. Whereas in our earlier work \cite{papathanasiout} we built the path integral by integrating products of free string propagators over the interaction points, here we treated it as a quantum field theory on the worldsheet. Given that free string propagators are the two-point functions of string field theory, we could call the former approach string field theory-based, and the latter approach worldsheet-based.

The key idea for treating string interactions in this framework, was to examine how the path integral is modified as we start removing links from the free worldsheet (\ref{lattice_action}). An essential ingredient for describing this departure, is the worldsheet correlation function $\Delta$ of two target space coordinates (\ref{wscorrelator_definition}). We consider as the main result of this paper, the determination of this quantity explicitly in Fourier mode space (\ref{wsprop_fourier}), and as a simple sum in coordinate space (\ref{neumann_open_Delta})-(\ref{wspropc}).

We then moved on to assess the efficiency of the worldsheet approach, by calculating the one-loop self-energy corrections for the closed string tachyon and graviton, and performing a comparison with the results of \cite{papathanasiout}. The self-energy corrections involve determinants of size $M$ and $K$ for the string field theory and worldsheet approach respectively, and hence the first one is more convenient for analyzing the $K\gg M$ (infrared) regime, whereas the second one for the $K\ll M$ (ultraviolet) regime.\footnote{We remind the reader that
$M$ is the spatial size of the lattice and $K$ is the temporal length of the
slit representing the loop.}

Indeed, with our current approach we were able to not only find the structure of the asymptotic expansion in $M$ of the self-energy summand, but also to calculate its coefficients exactly, for each value of $K$. In this manner, we were able to rigorously prove two important facts, for which we only had strong indications up to now. Namely, that the leading term in the expansion is the same for the tachyon and the graviton, and that the $\mathcal{O}(1/M^2)$ subleading term for the graviton is zero (i.e. the graviton is massless in the ultraviolet region). In contradistinction,  analyzing the $K\ll M$ regime in \cite{papathanasiout} had to rely on fits for both variables $M$ and $K$, which introduced larger numerical errors and made conclusions less definitive.

Apart from its calculational virtues, our new approach
adopts the point of view, implicit in our 
representation of the planar sum as a sum
over Ising spin variables (\ref{action_isinglike}),  
which is much closer to the treatment of 
more general lattice systems: Each string diagram is like a lattice state, and we build all states by gradually adding more and more `excitations', namely missing links, to the `vacuum', or free worldsheet. It would be very interesting to explore this point of view further, as it seems to suggest that string diagrams of different loop order but same excitation number may be similar to each other. Indeed, generalizing the considerations of sections (\ref{sec_longslit}) and (\ref{sec_longslit_graviton}) for arbitrary positions of the $K-1$ missing links $(m,l)=\{(k_1,j_1),\ldots,(k_{K-1},j_{K-1})\}$, yields again a $(K-1)$-dimensional determinant, this time with elements
\be
h_{ml,m^\prime l^\prime}=\delta_{m m^\prime}\delta_{m m^\prime}+\Delta_{(m+1)l,m^\prime l^\prime}-\Delta_{ml,m^\prime l^\prime}+\Delta_{ml,(m^\prime+1) l^\prime}-\Delta_{(m+1)l,(m^\prime+1) l^\prime}\,.
\ee
It may be more advantageous to organize the sum over diagrams not by 
loop order, as dictated by the conventional wisdom of 
string perturbation theory, but by the number of missing link `excitations'.

With this more efficient method now in place, a primary objective will be its application for the study of the one-loop self-energy corrections to the low-lying states of open string theory, which though more intricate, is of main interest because of its relation to large $N$ gauge theory. Once we have similarly established the compatibility of the lattice regularization with Lorentz invariance in this case as well, then the next natural step will be the numerical evaluation of the full path integral with the help of Monte Carlo methods.

In this respect, it will be very interesting to examine whether efficiency can be further improved by performing the sums in (\ref{neumann_open_Delta})-(\ref{wspropc}) analytically, in order to obtain explicit expressions for the worldsheet propagators in coordinate space as well. In the most probable scenario, that the summation of all bosonic string diagrams does not succeed in stabilizing the vacuum, we will of course be aiming to develop a similar treatment for the superstring as well.
\vskip12pt
\noindent{\bf Acknowledgements}:
This research was supported in part by the Department
of Energy under Grant No. DE-FG02-97ER-41029.

\appendix
\section{Normal Modes}
\label{normalmodes}
A string with $P^+=MaT_0$ is described at a fixed time by
$M$ coordinates $x_i$ or $y_i$, $i=1,\cdots M$. In this article we require
several normal mode decompositions depending on the boundary conditions.
\vskip12pt
\noindent Neumann Open String
\bea
x_i&=&{1\over\sqrt{M}}q_{0}+\sqrt{2\over M}\sum_{m=1}^{M-1}q_{om}
\cos{m\pi(i-1/2)\over M}\,,\\
q_0&=&\sqrt{1\over M}\sum_{i=1}^M x_i,\qquad\quad q_{om}=\sqrt{2\over M}\sum_i x_i\cos{m\pi(i-1/2)\over M}\,.
\label{normalNopen}
\eea
\vskip12pt
\noindent Dirichlet Open String
\bea
y_k&=&\sqrt{2\over M}\sum_{m=1}^{M-1}q_{Dm}
\sin{m\pi k\over M}\quad{\rm for}\quad k=1,\cdots,M-1,\qquad
y_M=q_{DM}\,,\\
q_{Dm}&=&\sqrt{2\over M}\sum_{k=1}^{M-1}y_k\sin{m\pi k\over M},\quad
0<m<M,\qquad
q_{DM}=y_M\,.
\label{normalDopen}
\eea
\vskip12pt
\noindent Closed String
\bea
x_k&=&{1\over\sqrt{M}}\sum_{m=0}^{M-1}
A_m\exp{2mk\pi i\over M},\qquad
A_m={1\over \sqrt{M}}\sum_k x_k
\exp{2(M-m)k\pi i\over M}\,.
\label{normalclosedexp}
\eea
This goes to the normal mode expansion with trigonometric functions with the
substitutions $A_m=A_{M-m}^*={(q_{cm}-iq_{sm})/\sqrt{2}}$,
with $0<m<{M/2}$, $A_0=q_0$, and $A_{M/2}=q_{cM/2}$ (if $M$ is even).
From this dictionary the non-zero  correlators are
\bea
\VEV{A_m A_{M-m}}&=&{1\over2}(\VEV{q_{cm}q_{cm}}+\VEV{q_{sm}q_{sm}})\qquad
m\neq 0,{M\over2}\,,\nonumber\\
\VEV{A_0A_0}&=&\VEV{q_0q_0},\qquad \VEV{A_{M/2}A_{M/2}}=\VEV{q_{M/2}q_{M/2}}\,.
\eea
\section{String Propagators}
\label{propagators}
\subsection{Neumann Open String Propagator}
\bea
\VEV{N+1,x^f|0,x^i}^{open}&=&{\cal D}^{open}(N+1)e^{iW_{open}}\,,\\
iW_{open}&=&-{T_0\over2}\bigg[{(q_{0,f}-q_{0,i})^2\over N+1}\nonumber\\
&&\hskip-.75in+\sum_{m=1}^{M-1}\sinh\lambda^o_m
\left((q_{m,i}^2+q_{m,f}^2)\coth(N+1)\lambda^o_m
-2{q_{m,i}q_{m,f}\over\sinh(N+1)\lambda^o_m}\right)\bigg]\,,\\
\lambda^o_m&=&2\sinh^{-1}\left(\sin{m\pi\over2M}\right)\,.
\eea
Where the $q_m$'s are the normal mode coordinates for the $x$'s.
The right side is the result of doing the integrations over
all the $x_i^j$ with $i=1,\cdots, M$ and $j=1,\cdots N$. The
propagator spans $N+1$ time steps and this result corresponds to
assigning half the potential energy $T_0\sum_{i=1}^{M-1}(x_{i+1}^j
-x_i^j)^2/2$ to time $j=0$ and half to $j=N+1$.
\subsection{Dirichlet Open String Propagator}
The Dirichlet open string propagator over a time of $K=N+1$ steps is
evaluated to be
\bea
\VEV{q^f,N+1|q^i,0}^{D}&=&{\cal D}^D(N+1)e^{iW^D}\,,
\eea
where
\bea
iW^D&=&-{T_0\over2}\left[\sum_{m=1}^{M}\left((q_{Dm}^{f2}+q_{Dm}^{i2})
\sinh\lambda_m^D\coth K\lambda_m^D
-2q_{Dm}^{f}q_{Dm}^{i}{\sinh\lambda_m^D\over
\sinh K\lambda_m^D}\right)\right]\\
{\cal D}^D(N+1)&=&\left({T_0\over2\pi}\right)^{M/2}
\prod_{m=1}^{M}\left[{\sinh(N+1)\lambda^D_m
\over\sinh\lambda^D_m}\right]^{-1/2},\\
\lambda^D_{M}&=&2\sinh^{-1}{1\over\sqrt{2}},\qquad \lambda^D_{m}
=\lambda^o_m=2\sinh^{-1}\sin{m\pi\over2M},\quad m=1,\cdots M-1\,.
\eea
We recall that the above expressions give the the result of integrating over
all the variables $y^j_i$, for $j=1,\cdots,N$, with half the potential
energy assigned to $j=0,N+1$, which is consistent with the
closure requirement.

\subsection{Closed String Propagator}
\bea
\VEV{N+1,x^f|0,x^i}^{closed}&=&{\cal D}^{closed}(N+1)e^{iW_{closed}}\,,\\
iW_{closed}&=&-{T_0\over2}\bigg[{(q_{0,f}-q_{0,i})^2\over N+1}\nonumber\\
&&\hskip-.75in+\sum_{m=1}^{M-1}\sinh\lambda^c_m
\left((q_{m,i}^2+q_{m,f}^2)\coth(N+1)\lambda^c_m
-2{q_{m,i}q_{m,f}\over\sinh(N+1)\lambda^c_m}\right)\bigg]\,,\\
\lambda^c_m&=&2\sinh^{-1}\left(\sin{m\pi\over M}\right)\,.
\eea
Where the $q_m$'s are the normal mode coordinates for the $x$'s.
When we divide the closed string normal modes into sine and cosine
modes, we arbitrarily call the $m>M/2$ modes sine modes and the
$m<M/2$ modes cosine modes. When $M$ is even, the $M/2$ mode
is not doubles.
The right side is the result of doing the integrations over
all the $x_i^j$ with $i=1,\cdots, M$ and $j=1,\cdots N$. The
propagator spans $N+1$ time steps and this result corresponds to
assigning half the potential energy $T_0\sum_{i=1}^{M}(x_{i+1}^j
-x_i^j)^2/2$ to time $j=0$ and half to $j=N+1$. In sums like these
it is understood that $x_{M+1}^j\equiv x_1^j$. Whenever we concatenate
at a time $j$ propagators with different numbers of missing links, we will
understand that we {\it add} terms $T_0(\Delta x)^2/4$ in the exponent
so that the
potential assigned to time $j$ is that of the system with the least number
of missing links. For example, the concatenation of an open string
propagator with a closed string propagator entails the addition
of $T_0(x_{M}^j-x_1^j)^2/4$ to the exponent.

\section{A Useful Euler-Maclaurin Expansion}\label{appx_EulerMcLaurin}
\begin{table}
\centering
\begin{tabular}{|c|c|}\hline
$n$&$I_n$ \\\hline
0&$1/2\simeq 0.5$\\
1&$-1/2 + 2/\pi\simeq 0.1366$\\
2&$-5/2 + 8/\pi\simeq0.04648$\\
3&$-25/2 + 118/(
  3 \pi)\simeq0.02019$\\
4&$-129/2 + 608/(3 \pi)\simeq0.01080$\\
5&$-681/2 + 16046/(
  15 \pi)\simeq0.006696$\\
6&$-3653/2 + 86072/(15 \pi)\simeq0.004568$\\\hline
\end{tabular}
\caption{Values of integral $I_n$ (\ref{EMintegral}), for $n=0,1,\ldots,6$.}
\label{tab:Euler_McLaurin_integral}
\end{table}

As we saw in in section \ref{sec_longslit} for the closed string propagator, when many links are missing, the leading term in the Euler-Maclaurin expansion of the elements of the corresponding determinant involves an integral of the form
\begin{eqnarray}
I_n&\equiv&\int_0^1dx{\sin\pi x\over\sqrt{1+\sin^2\pi x}}\left(\sin\pi x+\sqrt{1+\sin^2\pi x}\right)^{-2n}\label{EMintegral}\\
&=&\int_0^1dx{\sin \frac{\pi x}{2}\over\sqrt{1+\sin^2 \frac{\pi x}{2}}}\left(\sin \frac{\pi x}{2}+\sqrt{1+\sin^2 \frac{\pi x}{2}}\right)^{-2n}\nonumber\\
&=&\int_0^1dx{\sin \frac{\pi x}{2}\over\sqrt{1+\sin^2 \frac{\pi x}{2}}}\left(-\sin \frac{\pi x}{2}+\sqrt{1+\sin^2 \frac{\pi x}{2}}\right)^{2n}\nonumber\\
&=&\frac{2}{\pi}\int_0^1dz\frac{z \left(-z+\sqrt{1+z^2}\right)^{2n}}{\sqrt{1-z^2}\sqrt{1+z^2}}\nonumber\,.
\end{eqnarray}
We can evaluate this with the help of the identity \cite{determinants}
\be
\left(z+\sqrt{1+z^2}\right)^{2n}=\sum_{r=0}^n\lambda_{nr}z^{2r}+\sqrt{1+z^2}\sum_{r=1}^n\mu_{nr}z^{2r-1}\,,
\ee
where
\be
\lambda_{nr}=\frac{n}{n+r}\left(
\begin{array}{c}
n+r \\
2r
\end{array}\right)2^{2r}\,,\qquad \mu_{nr}=\frac{r \lambda_{nr}}{n}\,,
\ee
so that the integral can be rewritten as
\begin{eqnarray}
I_n&=&\frac{2}{\pi}\sum_{r=0}^n\lambda_{nr}\int_0^1dz\frac{z^{2r+1}}{\sqrt{1-z^2}\sqrt{1+z^2}}-\frac{2}{\pi}\sum_{r=1}^n\mu_{nr}\int_0^1dz\frac{z^{2r}}{\sqrt{1-z^2}}\nonumber\\
&=&\sum_{r=0}^n\lambda_{nr}\frac{\Gamma(\frac{1}{2}+\frac{r}{2})}{2\sqrt{\pi}\Gamma(1+\frac{r}{2})}-\sum_{r=1}^n\mu_{nr}\frac{\Gamma(\frac{1}{2}+r)}{\sqrt{\pi}\Gamma(1+r)}\,.
\end{eqnarray}
If desired, we can formally express these finite sums in terms of hypergeometric functions, for example
\be
\sum_{r=1}^n\mu_{nr}\frac{\Gamma(\frac{1}{2}+r)}{\sqrt{\pi}\Gamma(1+r)}=n\, _2F_1(1-n,1+n;2;-1)\,.
\ee
In any case, the sums can be readily evaluated for specific values of $n$, and for the reader's convenience we have tabulated the first few cases in table \ref{tab:Euler_McLaurin_integral}.

Summarizing, the sums that are relevant for the computation of the closed string propagator when many missing links are present, have an Euler-Maclaurin expansion of the form
\be
\frac{1}{M}\sum_{m=0}^{M-1}\sin\frac{\pi m}{M}{\left(\sin\frac{\pi m}{M}+\sqrt{1+\sin^2\frac{\pi m}{M}}\right)^{-2n}\over\sqrt{1+\sin^2\frac{\pi m}{M}}}=I_n-\frac{\pi}{6M^2}+\frac{(-1 + 3 n^2) \pi^3}{90 M^4}+\mathcal{O}
\left({\frac{1}{M^6}}\right)\,.
\ee


\end{document}